# Observation of Structure Evolution and Reaction Intermediates at the Gate-tunable Suspended Graphene/Electrolyte Interface

Ying Xu[1], You-Bo Ma[1], Feng Gu[1], Shan-Shan Yang[1], Chuan-Shan Tian[1,2*]

[1]Department of Physics, State Key Laboratory of Surface Physics and Key Laboratory of Micro- and Nano-Photonic Structures (MOE), Fudan University; Shanghai, 200433, China.

[2] Collaborative Innovation Center of Advanced Microstructures; Nanjing, 210093, China.

*Corresponding author. Email: cstian@fudan.edu.cn

**Abstract:** Graphene serves as an ideal platform to investigate the microscopic structure and reaction kinetics at the graphitic electrode interfaces. However, graphene is susceptible to various extrinsic factors, e.g. substrate, causing much confusion and controversy. Hereby we have obtained cm-sized substrate-free monolayer graphene suspended on electrolyte surface with gate tunability. Using sum-frequency spectroscopy, we have observed the structural evolution versus the gate voltage at the graphene/water interface. The Stern layer structure is invariant within the water electrolysis window, but undergoes drastic change when switching on the electrochemical reactions. The electrode is turned from hydrophobic to hydrophilic near the onset of hydrogen evolution reaction due to hydrogen adsorption. The large-size suspended pristine graphene offers a new platform to unravel the microscopic processes at the graphitic electrode interfaces.

Graphitic electrode is commonly used in electrochemical reactions due to its excellent in-plane conductivity, structural robustness and cost-efficiency [1,2]. It serves as prime electro-catalyst support as well as a layered intercalation matrix [2-4] with wide applications in energy conversion and storage, such as fuel cells, electrolysis hydrogen generation, and lithium-ion batteries [1,5,6]. Understanding these electrochemical processes requires molecular-level knowledge of the structure, and the intermediate and product species at the graphitic electrode interface during the reaction. However, such information remains limited thus far, hampered by the difficulties in probing the solid-liquid interface in a practical environment [7,8]. Over the past decades, chemical-sensitive optical spectroscopic techniques combined with cyclic-voltammetry (CV) have been developed to investigate buried electrochemical interfaces [9-11]. Unfortunately, most of them suffer from either poor surface sensitivity or strong attenuation when the probe beams pass through the electrode or electrolyte to access the interface [9-11]. On the other hand, being the two-dimensional building block of graphite, graphene shares similar chemical properties with graphite [1,2,12], while its unique physical and chemical properties offer more varieties and tunability for developing state-of-the-art graphitic devices [13-16]. For instance, the chemisorption of atomic hydrogen on a graphene sheet has been demonstrated as a promising route for efficient hydrogen storage at a low cost [6,17,18]. More importantly, with monolayer thickness, the graphene/electrolyte interface is easily accessible by various optical and electron probes through graphene. Thus, with the help of surface-specific techniques [13,14,19], such as sum-frequency spectroscopy, the graphene/electrolyte interface provides an ideal platform to investigate the graphitic electrode interface during electrochemical reactions.

Despite the apparent advantages mentioned above and intensive studies, our current knowledge of the molecular structure at the interface between graphene electrode and the aqueous solution remains elusive or controversial in literature [15,19-21]. The difficulty lies in the fact that the fingerprints of the interfacial species are vulnerable to many extrinsic factors, such as the hydrophobicity of the substrate used to support graphene, as well as surface charge due to defects or band bending at the interface. For instance, theoretical calculations predicted graphene is hydrophobic, and the dangling O-H bond of water molecules protruding at the graphene/water interface is expected[21,22]. However, earlier experiments could not observe its spectroscopic fingerprints [19,23]. This contradiction had been attributed to the substrate, because the hydrophobicity is strongly affected by the substrate, known as the wetting transparency of graphene [20,24]. Recently, Montenegro *et al* reported their observation of dangling O-D mode at $CaF_2$/graphene/$D_2O$ interface [15]. Nevertheless, their conclusion remains questionable as the voltage applied on graphene is well beyond the water electrolysis window which raises the concern about the gaseous $H_2$ layer accumulated at the interface [15]. Meanwhile, existence of the substrate can significantly change the interfacial potential due to its surface chemistry or band bending [19,25], which would strongly affect the hydrogen-bond network of the interfacial water molecules. Therefore, a clean and substrate-free graphene electrode is urgently needed to unravel the intrinsic structural evolution as well as adsorption and accumulation of the intermediate and product species at the interface during electrochemical reactions, e.g. chemisorption of atomic hydrogen on graphene during hydrogen evolution reaction (HER) [6,17,18]. In this work, we have obtained centimeter-sized substrate-free monolayer graphene (MLG) sheet on aqueous electrolyte surface using a newly developed transfer scheme for CVD graphene. At such MLG electrode/electrolyte interface, we have studied the molecular structure in the Stern layer as a function of the gate voltage, using sum-frequency vibrational spectroscopy (SFVS) combined with CV measurement.

The copper substrate of CVD graphene was etched away by electrolysis, as illustrated in Fig. 1(A). Then the electrolyte (0.3 M $CuSO_4$) solution in the cell was diluted by pure water repeatedly to reduce the concentration of the unwanted ions ($Cu^{2+}$, $SO_4^{2-}$) to be less than 1 μM, with the MLG sample kept suspended on the surface of solution in the cell during the whole process (see Sec. A in Supplementary Materials (SM)). Finally, supporting electrolyte was added to water, and here we used KCl solution (0.1 M). The MLG suspended on the solution is about 6 mm×10 mm in size, as shown in Fig. 1 (B). Optical microscope image (Fig. 1 (C)) shows that the CVD-grown graphene sheet is intact without folding or shrinkage in the mesoscopic scale. Furthermore, the ratio of D-mode and G-mode $I_D/I_G$ in Raman spectrum (Fig. 1 (D)) is about 1:8 and is essentially the same for graphene as-grown on copper substrate and graphene suspended on the solution, indicating that few lattice defects exist on graphene and that our fabrication process did not introduce additional defects [26,27]. These results manifest the premium quality of our free-standing MLG sample.

We next explored the electrical tunability of the graphene/electrolyte interface. As illustrated in Fig. 2 (A), thin Pt wires (20 μm in diameter) were put in contact with the MLG sheet, benefitting from the superior mechanical strength of graphene [28]. The graphene was used as the working electrode and a piece of Pt foil was immersed in the electrolyte as the counter electrode (Fig. 2 (A)), while Ag/AgCl in saturated KCl solution was used as the reference electrode. All of the voltages presented in this paper is referenced to Ag/AgCl (see Sec. A in SM). We first performed cyclic voltammetry (CV) measurement on the sample (scan rate 100 mV/s) and compared the result with that obtained on highly orientated pyrolytic graphite (HOPG) electrode (Fig. 2 (B)). The response of graphene is essentially the same as that of graphite. The water electrolysis window of our MLG sample is located between +1.0 V and -0.2 V. In this potential range, no chemical reaction occurs and a stable electric double layer (EDL) form upon applying voltage and the total surface charge density ($\sigma$) can be tuned continuously. Note that $\sigma$ is different from the carrier density (*n*) in graphene because the graphene is generally doped when in contact with aqueous solution. According to the quantum capacitance model of graphene [29] and the EDL model at the graphene/electrolyte interface [30] (also see section B in SM), $n$ and $\sigma$ can be readily deduced with known Dirac point and the potential of zero charge (pzc, where $\sigma = 0$) of the interface. To characterize the Dirac point, following the protocols in refs [26,31], we measured the frequency shift of Raman G mode and the in-plane resistance of graphene at different gate voltage $V_G$. As shown in Fig. 2 (C) and (D), the Fermi level is at the Dirac point when $V_G = -0.15$ V, as the resistance reaches maximum while frequency of G mode is at minimum. Pzc of the MLG/water interface was found to be $0.5 \pm 0.1$ V (see Sec. E in SM), which is close to that on HOPG basal plane reported in literature [32].

With this gate-tunable pristine MLG/electrolyte interface, we have studied the interfacial water structure using phase-sensitive SFVS (experiment details given in Sec. C in SM) [33]. Here, the stretching vibrations of water molecules (ranging from 3000 cm$^{-1}$ to 3800 cm$^{-1}$) at the interface are resonantly excited by the frequency-tunable IR laser pulses during the sum-frequency generation (SFG) process. The resonant feature of the 2$^{nd}$-order nonlinear response $\chi^{(2)}(\omega_{IR})$, obtained from the collected SF signal, can be used to derive the interfacial water structure in EDL. As a reference of a water interface, Fig. 3(A) shows the imaginary part of $\chi^{(2)}(\omega_{IR})$ spectrum at air/water interface [34], which directly characterizes the resonances. Being a hydrophobic interface, the spectrum generally consists of a broad continuum band from 3000 cm$^{-1}$ to 3500 cm$^{-1}$ and a narrow positive peak at 3700 cm$^{-1}$. The former is attributed to the hydrogen-bonded O-H vibrations, and the latter to the vibration of the dangling O-H bond protruding at the top-most layer as depicted in the inner plot of Fig. 3(A) [35,36]. With MLG placed on the water surface (air/MLG/water interface), the Im$\chi^{(2)}$ spectrum changes significantly as shown in Fig. 3(A). The spectrum recorded at $V_G = 0$ V is also composed of two main resonant features as well as a negative background (dashed line in Fig. 3(A)). Unambiguously, the dangling O-H band can now be observed at the pristine MLG/water interface, signifying the hydrophobic nature of graphene. The dangling OH mode is red-shifted by ~100 cm$^{-1}$ with broader linewidth versus the counterpart at the air/water interface resulting from the interaction between dangling O-H bond and graphene, in good agreement with theoretical prediction [21]. Such O-H mode persists when $V_G$ is tuned from 1.0 V to -0.2 V as evidenced in Fig. 3(B). It means the interface remains hydrophobic in the water electrolysis window. Meanwhile, the Im$\chi^{(2)}$ spectrum shows strong dependence on the gate voltage in the bonded O-H band between 3000-3500 cm$^{-1}$ and the negative background. In order to understand the evolution of the O-H spectrum and the background, we need to quantitatively analyze the origin of $\chi^{(2)}$ at this interface.

As illustrated in Fig. 3 (C), the total second-order nonlinear susceptibility ($\chi^{(2)}$) from the graphene/electrolyte interface contains three origins [37,38]:

$$\chi^{(2)} = \chi_s^{(2)} + \chi_g^{(2)} + \chi^{(3)}\Psi \tag{1}$$

Here, $\chi_s^{(2)}$, $\chi_g^{(2)}$ and $\chi^{(3)}\Psi$ corresponds to the SF response from the Stern layer, graphene and the diffuse layer, respectively. Firstly, the Stern layer is composed of water molecules and ions that are closely packed near the electrode, and its structure is directly related to the energy and charge transfer across the interface during electrochemical reactions [39,40]. Hence, $\chi_s^{(2)}$ that characterizes the microscopic structure of the Stern layer is what we are interested in the most. Secondly, the electric quadrupole and magnetic dipole responses of the free carriers in graphene account for the negative background ($\chi_g^{(2)}$) in the observed spectrum plotted in Fig. 3(B) [38,41]. Knowing the Fermi level $\mu$ of graphene (relative to the Dirac point), $\chi_g^{(2)}$ can be calculated using the theory developed by the pioneering works [38,41,42]. Finally, water molecules in the diffuse layer polarized by the screening electric field would generate SF signal proportional to $\chi^{(3)}\Psi$ (see ref [37] and section E in SM). Here, $\chi^{(3)}$ is the 3$^{rd}$-order responses of bulk water, and $\Psi$ is the modified electric potential difference across the diffuse layer [37]. It is sensitive to the charge density at the interface and could contribute significantly to the SF spectra in the bonded-OH band (3000 - 3500 cm$^{-1}$) given in Fig. 3(B).

In order to obtain Im$\chi_s^{(2)}$, we need to separate Im$\chi_g^{(2)}$ and Im$\chi^{(3)}\Psi$ from the measured Im$\chi^{(2)}$. We first evaluate Im$\chi_g^{(2)}$ by theoretical calculation using the formulism given in ref [38] and compare it with that measured at 3800 cm$^{-1}$, where the SF signals from $\chi_s^{(2)}$ and $\chi^{(3)}\Psi$ are off-resonance and negligibly small. As shown in Fig. 3(D), excellent agreement is achieved between the theoretical and experimental values as a function of the chemical potential $\mu$ of graphene. To the best of our knowledge, this is the first quantitative experimental verification of SF response versus $\mu$ on graphene. Having confirmed the theoretical results, we can obtain the Im$\chi_g^{(2)}$ spectrum between 3000 – 3800 cm$^{-1}$, as presented in Fig. 3(E). As expected, the calculated Im$\chi_g^{(2)}$ spectrum fits well with the background of the measured spectra in Fig. 3(B) at all gate voltages (see Section E in SM for details).

As for the contribution from the diffuse layer, based on the method established in our previous work, the Im$\chi^{(3)}$ spectrum of water has been already obtained [37] and $\Psi$ can be deduced knowing the surface charge density $\sigma$ and the ion concentration . Fig. 3(F) shows the calculated Im$\chi^{(3)}\Psi$ spectra at different $V_G$. Through comparison of the two sets of spectra in Fig. 3(B) and 3(F), the sign flip of the bonded O-H band in the measured Im$\chi^{(2)}$ as $V_G$ is tuned from +1.0V to -0.2V is originated from the Im$\chi^{(3)}\Psi$ contribution when the electric field direction reverses in the diffuse layer.

The Stern layer spectrum ($Im\chi_s^{(2)}$) can now be obtained by subtracting the contributions of graphene and the diffuse layer from the measured SF spectra using Eq. (1). A contour plot of the $Im\chi_s^{(2)}$ spectrum at different $V_G$ is presented in Fig. 3(G), with a few representative curves given in Fig. 4(A). Surprisingly, although the total SF spectrum (see Fig. 3(B)) changes significantly with $V_G$, the Stern layer spectrum is essentially independent of $V_G$ between +0.7 V and 0.0 V, except for shift of the dangling OH band. It means the water structure close to graphene is hardly altered by the gate voltage within the water electrolysis window. The frequency shift of the dangling OH mode is the result of the vibrational Stark effect [43,44]. In Fig. 4(B), we fitted the frequency shift $\Delta \nu$ versus the electric field $E_s$ (positive if pointing towards graphene, see Section B in SM) in the Stern layer using the linear Stark effect:

$$\Delta \nu = -\alpha \cdot E_s \qquad (2)$$

Here $\alpha$ is the Stark tuning rate. From the linear fit in Fig. 4 (A), $\alpha = 7 \pm 2$ cm$^{-1}$/MV•cm$^{-1}$, close to the theoretical value of $\alpha = 4.8$ cm$^{-1}$/MV•cm$^{-1}$ for water clusters in literature [43].

Knowing the Stern layer is invariant within the water electrolysis window, we further explored the microscopic structure at the pristine graphene/electrolyte interface by switching on the electrochemical reactions. As shown in Fig. 4(A), when the gate voltage is near the onset of electrolysis of water, namely at $V_G$ = 1.0 V and -0.2 V, the bonded O-H band starts to change drastically. This indicates that the hydrogen-bond network has been strongly altered upon starting the chemical reactions. If we further lower the gate voltage down to -0.4 V, which is slightly beyond the threshold of HER, the dangling O-H mode disappears completely (Fig. 4(C)). The disappearance of the dangling O-H manifests that topmost layer of the interfacial water is strongly disturbed. In particular, the electrode has been turned from hydrophobic to hydrophilic [45]. Meanwhile, the CV curves in Fig. 4(D) measured immediately after holding $V_G$ at -0.4 V for

~4 hours show a peak at 0.3 V, and the peak current reduces after each loop. The peak at 0.3V indicates an irreversible desorption process of the products on graphene. Atomic H or molecular $H_2$ may be produced at the interface when $V_G$ = -0.4 V. However, gaseous $H_2$ molecules at the interface should favor existence of the dangling O-H bond in analogy to the air/water interface [15], rather than suppressing of the dangling O-H. Thus, the change of hydrophobicity of the electrode should be attributed to atomic H adsorbed on graphene prohibits dangling-like O-H at the interface, as illustrated in Fig. 4 (E). We note that in a recent published work, He *et al* also observed the signature of chemisorbed hydrogen atom on graphene at the onset of HER [18].

In summary, our work demonstrated that substrate-free pristine graphene is essential for unraveling the intrinsic microscopic structure at the graphene electrode interface. The suspended graphene can be easily decorated with electro-catalysts, e.g. Pt and Au nanoparticles [46,47], to promote the electrochemical reactions. Thus, the large-size substrate-free MLG with superior mechanical strength and electrical tunability provides us an ideal platform for investigation of the interfacial species and their reaction kinetics at the graphitic electrode/electrolyte interface in general.

**Acknowledgement:** We thank discussions with Carbon Six. Co for the CVD graphene growth and Dr. Yu-Dan Su for the analysis of data. CST acknowledges support from the National Natural Science Foundation of China Grants (No. 12125403 and No. 11874123) and the National Key Research and Development Program of China (No. 2021YFA1400503 and No. 2021YFA1400202).

**Author Contributions** C.S.T conceived the project and designed the experiments. Y.X, SS. Y, YB. M and F. G and performed experimental data collection. C.S.T and Y. X wrote the manuscript.

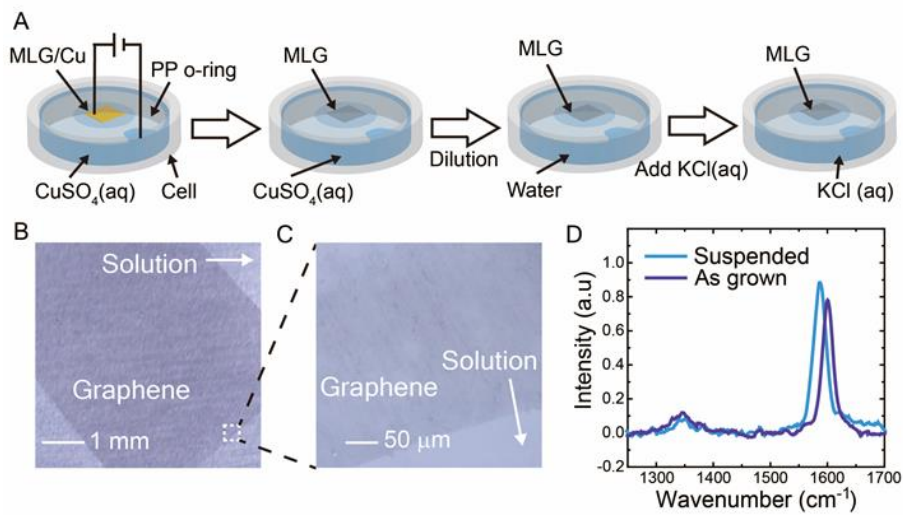

**Fig. 1.** Suspended MLG sample on water. **(A).** Fabrication process of the suspended MLG. The MLG on copper foil was put on the surface of $CuSO_4$ solution and a polypropylene (PP) O-ring was used to confine the sample in the central area of the cell. After etching copper substrate by electrolysis, the remaining electrolyte was diluted repeatedly with pure water, and then the supporting electrolyte KCl was added to reach 0.1 M. **(B).** Photograph of the MLG sample suspended on the electrolyte surface. **(C).** Enlarged microscope image of the sample in the white dashed box in **(B)**. **(D).** Raman spectrum of the suspended MLG sample compared to the graphene as-grown on the copper substrate.

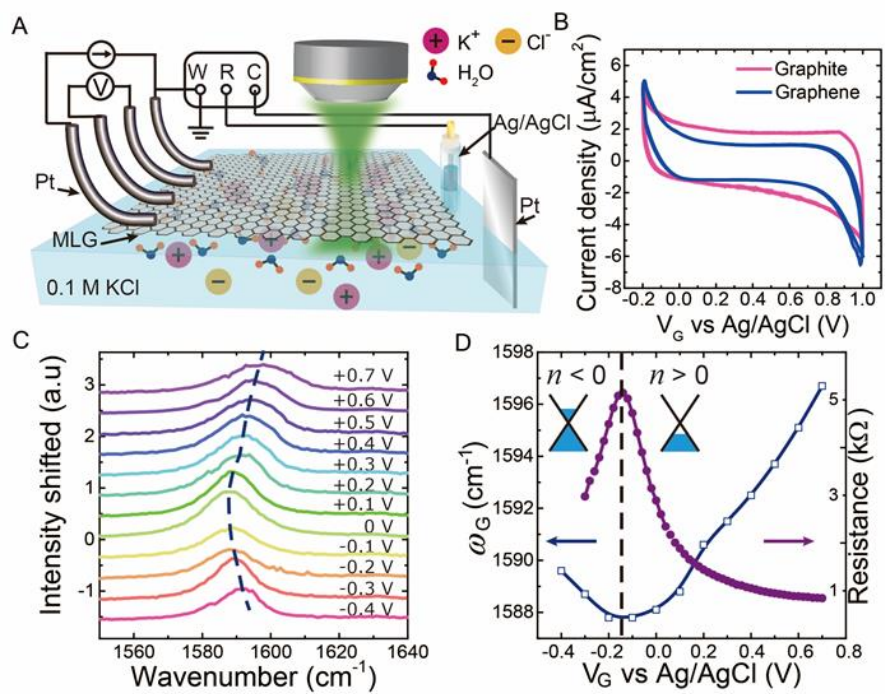

**Fig. 2. Gate tunability of the suspended MLG. (A).** Illustration of the gated MLG. Electronic circuit is sketched in the figure. W, R and C refers to working electrode, reference electrode and counter electrode, respectively. The four Pt wires (~ 20 μm) are attached to the graphene, to perform four-probed resistance measurements and to tune the gate voltage ($V_G$). Raman scattering photons are excited and collected through the same objective. **(B).** CV scans on graphene and graphite. **(C).** Raman spectra of the MLG sample at different $V_G$. Dashed curve marks the frequency shift of the G-mode. **(D).** Shift of the G-mode (blue dots) and the resistance (purple dots) of MLG suspended on water versus gate voltage $V_G$. Curves serve as eye-guides. Inset illustrates the Dirac cone and Fermi level of graphene.

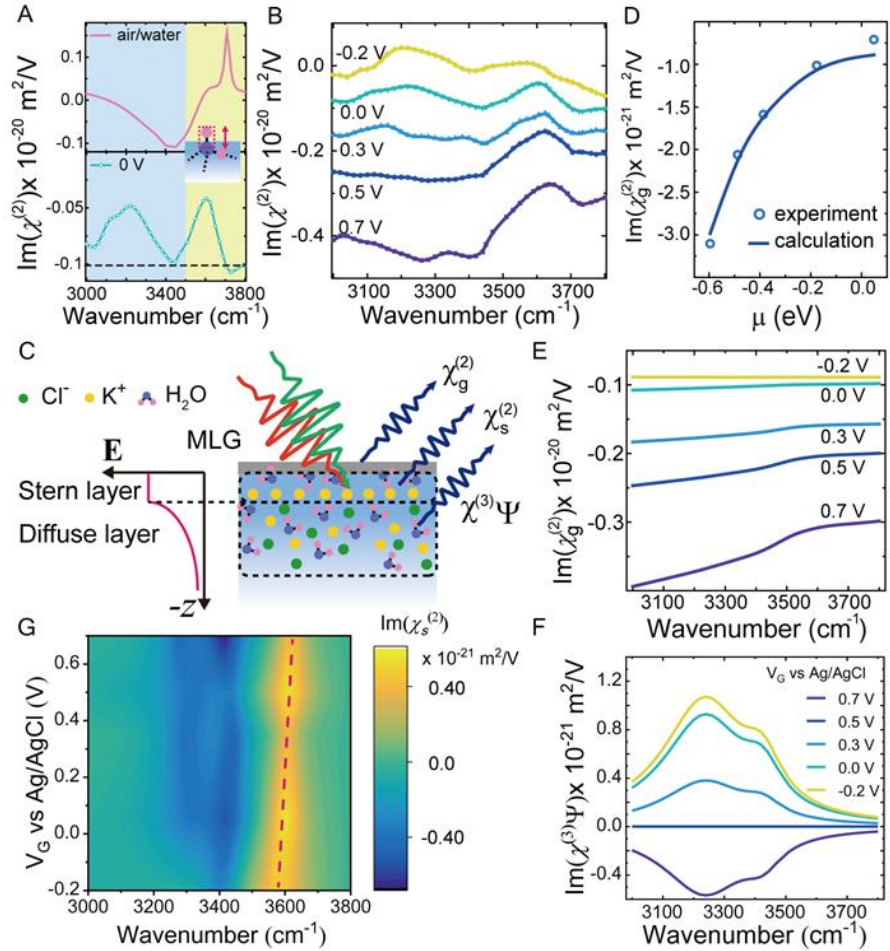

**Fig. 3. In situ SFVS spectra of the graphene/electrolyte interface. (A).** Comparison of Im$\chi^{(2)}$ spectrum at air/water interface and air/MLG/water interface. Inner plot illustrates the dangling O-H bond of water molecule protruding at the hydrophobic interface. **(B).** Im$\chi^{(2)}$ spectra at different $V_G$. **(C).** Illustration of the graphene/electrolyte interface with three origins contributing to the SF spectrum. The electric field distribution in the EDL is also sketched. **(D).** Theoretical and experimental data of Im$\chi_g^{(2)}$ at 3800 cm$^{-1}$ versus chemical potential μ of graphene. **(E).** Calculated SF spectra of graphene (Im$\chi_g^{(2)}$) at different $V_G$. Fresnel factors have been taken into account (see SM for details). (F). Calculated SF spectra from the diffuse layer (Im$\chi^{(3)}\Psi$) at different VG. **(G).** Contour plot of the Stern layer spectra (Im$\chi_s^{(2)}$) at different $V_G$. Red dashed line illustrates the frequency shift of the dangling O-H mode.

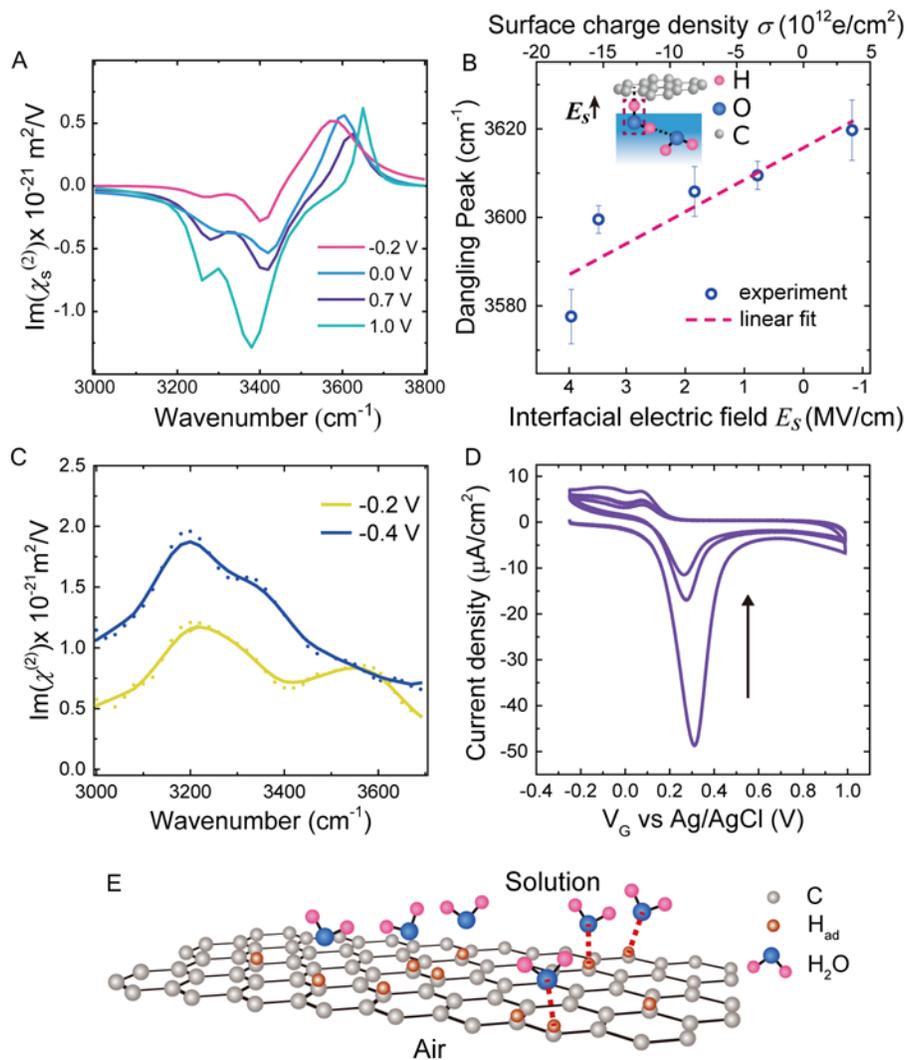

**Fig. 4. SF spectrum and CV curves near the onset of chemical reactions. (A)** Typical $\text{Im}\chi_s^{(2)}$ spectra from the Stern layer at different $V_G$. **(B).** Linear fit of the peak wavenumber of the dangling O-H mode. The interfacial electric field $E_s$ would induce Stark shift of the dangling O-H mode (marked in the red dashed box) as illustrated in the inset. Black arrow: direction of positive $E_s$. **(C).** Comparison of $\text{Im}\chi^{(2)}$ (after subtraction of the graphene background) at $V_G = -0.4$ V and $-0.2$ V. Dots are experimental data and solid curves serve as eye-guides. **(D).** CV scans immediately after holding $V_G$ at $-0.4$ V for ~4 hours. The black arrow indicates that the peak amplitude decreases after each loop. **(E).** Illustration of the adsorbed H atoms ($H_{ad}$) on graphene, which would modify the hydrophobicity of graphene. The red dashed lines represent interactions between $H_{ad}$ and water molecules, favoring the shown orientation of water molecules at the interface.